\documentclass{sf2a-conf2011}
\usepackage{graphicx}
\usepackage{hyperref}
\usepackage[]{natbib}  
\usepackage[cyr]{aeguill}
\usepackage{epstopdf}

\def\BibTeX{{\rm B\kern-.05em{\sc i\kern-.025em b}\kern-.08em
    T\kern-.1667em\lower.7ex\hbox{E}\kern-.125emX}}
\bibpunct{(}{)}{;}{a}{}{,}  

\begin{document}

\TitreGlobal{SF2A 2011}


\title{Stellar activity cycles and asteroseismology}

\runningtitle{Stellar activity cycles and asteroseismology}

\author{D. Salabert}\address{ Universit\'e de Nice Sophia-Antipolis, CNRS UMR 6202, Observatoire de la C\^ote d'Azur, BP 4229, 06304 Nice Cedex 4, France}

\setcounter{page}{237}

\index{Salabert, D.}


\maketitle

\begin{abstract}
The success of helioseismology is due to its capability to accurately measure the p-mode parameters of the solar eigenmode spectrum, which allow us to infer unique information about the internal structure and dynamics of the Sun from its surface all the way down to the core. It has contributed greatly to a clearer understanding of the Sun and provided insights into the complex solar magnetism, by means for instance of the variability of the characteristics of the p-mode spectrum. Indeed, variations in the mean strength of the solar magnetic field lead to significant shifts in the frequencies of even the lowest-degree p modes with high levels of correlation with solar surface activity proxies. These frequency shifts are explained to arise from structural changes in the outer layers of the Sun during the 11-year activity cycle, which is understood to be driven by a dynamo process. However, clear differences between p-mode frequencies and solar surface activity during the unusually extended minimum of cycle 23  were observed. The origin of the p-mode variability is thus far from being properly understood and a better comprehension of its relationship with solar and stellar activity cycles will help us in our understanding of the dynamo processes.
Spectroscopic measurements of Ca H and K emission lines revealed magnetic activity variations in a large sample of solar-type  stars with timescales ranging from 2.5 and 25 years. This broad range of cycle periods is thought to reflect differences in the rotational properties and the depths of the surface convection zones with various masses and ages. However, spectroscopic measurements are only good proxies of surface magnetic fields.
The recent discovery of variations with magnetic activity in the p-mode oscillation frequencies of the solar-like star HD 49933 observed by CoRoT, with a frequency dependence comparable in shape to the one observed in the Sun, opens a new era in the study of the physical phenomena involved in the dynamo processes. Current and future asteroseismic observations  will contribute to probe stellar cycles in a wide variety of solar-type stars.
\end{abstract}



\section{Introduction}
The Sun is a variable star with an 11-year cyclic variation of its magnetic activity. Signatures of its variability can be seen in several activity proxies, the most famous being the number of spots on the surface of the Sun. Other indices such as the 10.7-cm radio flux or the irradiance for instance also show this 11-year periodicity. The 11-year magnetic cycle is theoretically understood to be driven by a magnetic dynamo process located at the bottom of the convective zone \citep{svalgaard05,dikpati06}. Attempts have been made to provide prediction of the cycle properties based on dynamo models but these are not conclusive yet. Indeed, different conclusions were obtained for the new cycle 24, for which these models  had not predicted the long and deep minimum of cycle 23.

Although spots on solar-type stars cannot be directly observed, stellar magnetic cycles have been already reported. Indeed, the associated areas of concentrated magnetic field produce strong emission in the Ca II H and K spectral lines, which were proven to be a good proxy of surface magnetic fields \citep{Leighton59}.
Surveys from the Mount Wilson and Lowell Observatories (northern-hemisphere targets) along with long-term monitoring campaigns with SMARTS 1.5-m telescope at CTIO (southern-hemisphere targets) have revealed that many solar-type stars exhibit long-term cyclic variations in their Ca II H and K emission lines, analogous to the solar variations \citep{wilson78,baliunas95,metcalfe09}. The complete sample includes cycle periods covering a range between 2.5 to 25~years. Recently, \citet{metcalfe10} discovered an even shorter activity cycle of 1.6~years in  the exoplanet hosting F8V  star HD~17051.
Moreover, it has been observed that the periods of the activity cycles  increase proportional to the stellar rotational periods along two distinct paths in main-sequence stars: the relatively young, active sequence, and the older, less active sequence \citep{Saar99,BV07}. Relations between mean level of magnetic activity, rotation rate, and periods of the observed activity cycle generally support a dynamo interpretation.
However, spectroscopic measurements are only good proxies of surface magnetic fields.

The recent discovery, using asteroseismic observations, of variations with magnetic activity in the p-mode oscillation frequencies of the solar-like F5V star HD~49933 \citep{garcia10}, with a frequency dependence comparable in shape to the one observed in the Sun \citep{salabert11}, opens a new era in the study of the physical phenomena involved in the dynamo processes. Indeed, through the precise measurements of oscillation parameters, the seismic observables provide unique information about the interior of the stars and for the determination of key parameters for the study of stellar activity, such as the depth of the convection zone, the characteristic evolution time of the granulation, the differential rotation, or the sound-speed as a function of the star's radius \citep[see the preliminary work on HD~49933 by][]{ceillier11}. All these new observables will impose new constraints to the theory \citep[e.g.][]{chaplin07a,metcalfe07}.
 Long, high-quality photometric observations such as the ones collected by the space-based  Convection, Rotation, and planetary Transits \citep[CoRoT,][]{baglin06} and {\it Kepler} \citep{koch10} missions will contribute to probe stellar cycles in a wide variety of solar-type stars.

\section{Variability of the oscillation parameters with solar magnetic activity}
The success of helioseismology is due to its capability to accurately measure the p-mode parameters of the solar eigenmode spectrum, which allow us to infer unique information about the internal structure and dynamics of the Sun from its surface all the way down to the core. 
It has contributed greatly to a clearer understanding of the Sun and provided insights into the complex solar magnetism, by means for instance of the variability of the characteristics of the p-mode spectrum.
Evidence of p-mode frequency changes with solar activity, first revealed by \citet{wood85}, were established by \citet{palle89} with the analysis of helioseismic observations spanning the complete solar cycle~21 (1977--1988). As longer, higher quality, and continuous helioseismic observations became available, the solar p-mode frequencies proved to be very sensitive to the solar surface activity \citep[e.g.,][and references therein]{anguera92,regulo94,chano01,gelly02,salabert04,chaplin07b} with high levels of correlation with solar surface activity proxies. The low-degree, p-mode frequencies change by about $0.4 \mu$Hz between the minimum and the maximum of the solar cycle with correlation levels with surface activity proxies higher than 0.9. Moreover, \citet{howe02} showed that the p-mode frequencies are shifted in presence of surface magnetic activity, varying with close temporal and spatial distributions. However, clear differences were observed between the frequency shifts and the surface activity of the Sun during the unusually extended and deep minimum of cycle 23 \citep{howe09,salabert09,broomhall09,jain11}.  Although the form and the dependence of the shifts indicate a near-surface phenomenon explained to arise from changes in the outer layers of the Sun, the origin of the frequency shifts is far from being properly understood. Moreover, a quasi-biennial signal in the solar frequencies was recently observed by \citet{fletcher10}, indicating the possible action of a second dynamo seated near the bottom of the layer extending 5\% below the solar surface.

The other p-mode oscillation parameters such as the amplitude, the lifetime, and the asymmetry, were also observed to vary with solar activity \citep{chaplin00,komm02,salabert03,salabert06,chano07}. For example, the mode amplitudes decrease  with increasing solar activity by about 40\% from minimum to maximum of the 11-year cycle. The temporal variations between frequency and amplitude are therefore anticorrelated.

\section{Signatures of stellar magnetic activity using asteroseismic observations}
The solar-like F5V star HD~49933 observed by the CoRoT satellite \citep{appourchaux08,benomar09} is the first star after the Sun for which variability of the p-mode parameters with magnetic activity was observed \citep{garcia10}. Indeed, the p-mode frequencies and amplitudes vary with time with a clear anti-correlation  between both parameters as observed in the Sun \citep[e.g.][]{salabert03}. These temporal variations suggest a cycle period of at least 120 days. 
Preliminary spectroscopic measurements with of the Ca II H and K emission lines had shown that HD~49933 is an active star, with a Mount Wilson S-index of 0.3. Follow-up long-term monitoring confirms the existence of a short activity cycle (T. Metcalfe, private communication, 2011). Incoming observations (both asteroseismic and spectroscopic) will help us to determine more accurately the period and the properties of this magnetic activity cycle.
\citet{salabert11} showed that the frequency shifts measured in HD~49933 present a frequency dependence with a clear increase with frequency, reaching a maximal shift of about $2 \mu$Hz around $2100 \mu$Hz, which shows a similar pattern as in the Sun. That indicates then the presence of similar physical phenomena driving the frequency shifts of the oscillation modes as the ones taking place in the Sun, which are understood to reflect structural changes in and just below the photosphere with stellar activity if we suppose similar mechanisms as in the Sun \citep[e.g.,][]{goldreich91}. 
However, the frequency shift measured in HD~49933 is at least five times larger than in the Sun, which reaches about $0.5 \mu$Hz at $3700 \mu$Hz between the maximum and the minimum of the 11-year solar cycle \citep[e.g.][]{salabert04}. 
This observation supports the scaling proposed by \citet{metcalfe07}, who predicted that stars hotter and more evolved than the Sun (like HD~49933) should have larger frequency shifts than in the Sun.
Preliminary structure models of HD~49933 were computed by \citet{ceillier11} in order to study the effects of sound-speed perturbations in the near surface layers on the p-mode frequencies. This will provide for instance insights on the properties of the convective zone of HD~49933 and on the depth at which the magnetic field perturbs the modes.

Three other solar-like stars observed by CoRoT have been studied, for which spectropolarimetric measurements from the NARVAL instrument \citep{auriere03} located at the Pic du Midi Observatory are available: HD~181420 \citep{barban09}, HD~49385 \citep{deheuvels10}, and HD~52265 \citep{ballot11}. Although HD~181420 and HD~49385 seem to be in a quiescence state, HD~52265 shows small temporal variations of the p-mode parameters, suggesting a modest increase of magnetic activity, also indicated by the spectroscopic observations \citep{mathur11}.

\section{Conclusions}
Asteroseismology provides not only invaluable information to infer the structure and the dynamics of the interior of the stars but also key parameters for the study of stellar magnetic activity cycles in a wide variety of solar-type stars  imposing new observational constraints to the theory. Studies of stellar activity cycles will bring important inferences for the modeling of dynamo processes and will put the Sun and its 11-year activity cycle into context. The solar-like F star, HD~49933, is the first star after the Sun for which variations of the oscillation parameters with magnetic activity have been observed.  Although these changes follow analogous patterns as in the Sun, they suggest a short cycle period, which is confirmed by spectroscopic observations. They also support that stars hotter and more evolved than the Sun should have larger frequency shifts. This is important for CoRoT and {\it Kepler} when searching for similar frequency shifts as it suggests that the shifts should be easier to detect in the F stars. Also, as short activity cycles seem to be more common than expected, the CoRoT and {\it Kepler} (and future) missions should potentially be able to observe a large number of full cycles. Such studies will also provide inputs to explore  the impacts of magnetic activity on possible planets hosted by these stars.

\begin{acknowledgements}
The author acknowledges the financial support  from the Centre National d'Etudes Spatiales (CNES).
\end{acknowledgements}


%
\end{document}